\documentclass[11pt]{amsart}

\usepackage{amsmath}
\usepackage{amsfonts}
\usepackage{amsthm}

\newtheorem{thm}{Theorem}[section]

\newtheorem{conj}[thm]{Conjecture}

\theoremstyle{definition}

\newtheorem{question}[thm]{Question}

\begin{document}
\title[Turing's Landscape]{Turing's Landscape: decidability, computability and complexity in string theory}
\author{Abhijnan Rej}
\thanks{{\tt abhijnan.rej@gmail.com}}
\thanks{Prepared for the 2nd FQXI Essay Contest \emph{What is Ultimately Possible in Physics?}. This work was partially carried out during a stay at the Center for Contemporary Studies, Indian Institute of Science, Bangalore in August 2009. I thank the Center for its hospitality and interest.}
\begin{abstract}
I argue that questions of algorithmic decidability, computability and complexity should play a larger role in deciding the ``ultimate" theoretical description of the Landscape of string vacua. More specifically, I examine the notion of the average rank of the (unification) gauge group in the Landscape, the explicit construction of Ricci-flat metrics on Calabi-Yau manifolds as well as the computability of fundamental periods to show that undecidability questions are far more pervasive than that described in the work of Denef and Douglas. 
\end{abstract}
\maketitle
\begin{verse}
\small{Three rocks, a few burnt pines, a lone chapel\\
and farther above\\
the same landscape repeated starts again:\\
-- George Seferis, \emph{Mythistorema}}
\end{verse}
\section{Introduction}
Fundamental physical theory finds itself, circa 2009, in a rather peculiar position. On one hand we have the Standard Model (SM) of particle physics with its three generations of quarks and leptons and the force mediating gauge bosons. The SM has been verified by all experiments so far with the single exception of the Higgs Boson and it is widely anticipated that it would be detected in the first runs of the LHC in the next couple of years. We also have general relativity (GR) which works remarkably well as a theory of gravity, atleast at the intermediate cosmological scales. However, when we try to unify quantum theory with GR (let alone the quantum SM with GR!) we reach a road block-- for one, we do not have a \emph{proof} whether this is \emph{fundamentally impossible} or not. A popular claim is that a supersymmetric theory of strings and membranes in a 10-dimensional space provides \emph{the} consistent unification of quantum field theory with GR incorporating the SM as a low energy effective theory.
\par
The problems with this approach has been highlighted and much debated in the last couple of years (cf. \cite{smolin}). The scientific crux of the argument against superstrings/M-theory relies on one major problem-- instead of predicting a unique vacuum state which would, putatively, describle our universe, strings/M-theory admits a very large number of vacua. This collection of vacua, in analogy to problems in evolutionary biology, has been termed the string Landscape. 
\par
Much ink has been spent on merits and demerits of the string Landscape. Opponents have decried it as an example of how wrong a scientific theory can be in terms of not being falsifiable. Proponents have argued that, eventually, by studying the statistics of the Landscape-- i.e. of the configuration space $\mathcal{C}$ of string vacua, $|\mathcal{C}| \sim 10^{500}$-- coupled to one version of the anthropic principle or the other, we would be able to identify a unique state that describes our universe. In fact, some cosmologists (for example Vilenkin \cite{vilenkin}) have argued that, along with the KKLT moduli stabilization mechanism and eternal inflation, the existence of the string Landscape points to a \emph{multiverse} where each causally disconnected part is a universe of its own (literally!) separated from another by domain walls and each with its own set of free parameters, unification gauge group etc.
\par
The goal of this Essay is to examine some computation-theoretic aspects of this multiverse picture to draw the conclusion that \emph{it may be fundamentally impossible to seperate one point in the Landscape from the other}. I will do so by examining some decidability problems associated to the choice of an average unification gauge group in the multiverse. I will claim that the moduli space of metrics on a Calabi-Yau manifold has a fractal structure and argue, following the work of Nabutovsky and Weinberger, that there are several computability issues associated to the explicit construction of a Ricci-flat metric on a Calabi-Yau manifold. I will also argue that the problem of deciding whether two points on the Landscape have the same fundamental period may also be computationally intractable.
\par
At the outset, I must mention the predecessor to this line of thought-- Denef and Douglas, in an influential paper a couple of years ago, had shown that the problem of matching the observed (small) value of the cosmological constant in the Bousso-Polchinski model was NP-complete \cite{denefdouglas}. In the concluding sections of \cite{denefdouglas}, they make several brief remarks which form the germ of this work.
\par
I shall make extensive use of certain results in pure mathematics (mostly in algebraic and differential geometry and topology). This means, regrettably, that parts of this essay may appear quite technical. To minimize demands on the part of my readers I have kept the mathematical arguments at the bare neccessary minimum. (In a paper currently under preparation \cite{rej}, I shall furnish the full mathematical details for the arguments made here.)

\section{Decidability and gauge groups}
One of the greatest discoveries in twentieth century physics has been that our material universe is best described in terms of local and global gauge symmetries. The Standard Model has, as input data, three Lie gauge groups: the abelian gauge group $U(1)$ describing electromagnetism and the nonabelian gauge groups $SU(2)$ and $SU(3)$ describing weak and strong interactions respectively. Through a phenomenom of gauge mixing between $SU(2)$ and $U(1)$, the gauge symmetry underlying the SM is
\begin{equation}\label{eq:SM} 
G_{SM} := SU(3) \times SU(2) \times U(1). 
\end{equation}
(It is customary to consider the quotient of the r.h.s of (\ref{eq:SM}) by $\mathbb{Z}/{6 \mathbb{Z}}$.) All elementary particles are then described in terms of representations of the corresponding Lie algebras.
\par
By running the coupling constants to a sufficiently high energy ($\sim 10^{16}$ GeV) and some fine-tuning, one expects that the three fundamental forces of nature would be unified in the sense that $G_{SM}$ would be a subgroup of a larger unification group $G$. There are some basic representation-theoretic restrictions on what $G$ can or can not be based on the fact that we should at lower energies see $G_{SM}$ of the form (\ref{eq:SM}). These are (cf. \cite{distlergaribaldi}) that, for $E$ a fixed real Lie group, 
\begin{enumerate}
\item{$G$ should be a subgroup of $E$ which is connected, reductive and compact and centralizing the Lorentz group $\mathop{\rm SL}(2,\mathbb{C})$,}
\item{Chirality conditions on the $G_{SM}$: $V_{2,1}$ is a complex representation of $G$ ($V_{m,n}$ denotes complex representation of $G \times \mathbb{C}$), and} 
\item{No exotic higher spin particles: $V_{m,n} = 0$ if $m+n > 4$.}
\end{enumerate}
Popular choices for $G$ has been Georgi--Glashow $SU(5)$ (ruled out by proton decay experiments), $\mathop{\rm Spin}(10)$, $E_6$ and the Pati--Salam group ${(\mathop{\rm Spin}(6) \times \mathop{\rm Spin}(4))}/{\mathbb{Z}/{2\mathbb{Z}}}$.
\par
Of course, these requirements on $G$ are for a very distinguished point on the Landscape, namely the universe we live in! In principle, different points on the Landscape could have wildly different $G$. Of course, we would \emph{like} to show that ``almost all" points in the configuration space have $G$ with a subgroup $G_{SM}$ of the form (\ref{eq:SM}) (the ``naturalness condition") but diversity arguments (akin to those in evolutionary biology) forces us to consider very general Lie groups with diverse subgroups.
\par
The central parameter for statistical analysis of gauge groups in the Landscape is the average \emph{rank} of the gauge group (the average taken over the entire configuration space $\mathcal{C}$ with respect to a suitable measure). This average rank is then expressed in terms of the number of complex moduli of the compactified space and the configuration of D-branes wrapping it (in terms of flux). 
\par
More precisely, following \cite{kumarwells}\footnote{Similiar analysis was also done by Blumenhagen et. al. \cite{blumenhagen}}, for $X$ a CY 4-fold such that the orientifold limit of the F-theory compactified on $X$ is of type IIB on the orientifold $Y$, the average rank of a D3-brane gauge group is \cite{kumarwells}
\begin{equation}\label{eq:gaugprk}
\langle R_{D3} \rangle = \frac{L_{*}}{2n+3},
\end{equation}
where $n$ is the number of complex structure moduli of $Y$ and 
$$L_{*} = N_{D3} + \int F^{RR} \wedge H^{NS}.$$
($N_{D3}$ is the net D3-brane charge.) We additionally require the tadpole cancellation condition $L_{*} - N_{D3} = \frac{\chi(X)}{24}$ where $\chi(X)$ is the Euler characteristic of $X$. In presence of a small cosmological constant $\Lambda_{*}$, the gauge group rank average is not significantly different from (\ref{eq:gaugprk}):
$$\langle R_{D3} \rangle_{\Lambda_{*}} = \frac{L_{*}}{2n+2}.$$
A notable fact in (\ref{eq:gaugprk}) is the absence of any parameter that depends on the explicit structure of the CY 3-fold. It is shown in \cite{kumarwells} that for $Y$ the standard orientifold $\mathbb{T}^6/{\mathbb{Z}_2}$ with symmetric flux, $n=1$ and $L_{*} = 16$, so $\langle R_{D3} \rangle = \frac{16}{5}$ which is close to the SM gauge group rank of $4$. Of course, in presence of a small c.c. for the standard orientifold, $\langle R_{D3} \rangle_{\Lambda_{*}} = \frac{16}{4} = 4$ which is exactly the SM gauge group rank. It is also shown that the fraction of all SUSY vacua that have gauge group rank $R$ (of, possibly, the unification gauge group) above the SM gauge group rank $R_{SM}$ is
\begin{equation}
\eta \sim \exp\Big(-\frac{R_{SM}}{\langle R \rangle}\Big)
\end{equation}
in the large $n$ limit. In related analysis Gmeiner et. al  \cite{gmeiner} estimate that the frequency of occurence of minimally supersymmetric standard model in the Landscape (with supersymmetric intersecting D-branes on an toriodal orientifold background) is of the order $10^{-9}$.
\par
Let us fix a point in $\mathcal{C}$ and write the gauge group rank of this point as ${\bf r}$ and the corresponding gauge (Lie) group as ${\bf G}$. Let us furthermore imagine that there exists a sequence of Lie groups $G_i$ with rank of $G_i = \alpha_i$ for $i \in I$ a finite index set such that \emph{either} each individual $G_i$ is a subgroup of ${\bf G}$ satisfying condition (1) above \emph{or} a product of $G_i$ is a subgroup of ${\bf G}$ satisfying (the rather mild!) condition (1). Other than satifying the condition (1), each $G_i$ or their product could be any Lie group and of any real dimension; the mechanics of flux compactification does not in principle prohibit this.
\par
The question at hand is
\begin{question}\label{q:gauge}
\emph{Given the rank ${\bf r}$ of ${\bf G}$ and the ranks of $G_i$ being $\alpha_i$, can we find a subsequence $G_k$, $1 \leq k < n$, $n = |I|$ such that $\alpha_1 + \alpha_2 + \ldots + \alpha_k = {\bf r}$?}
\end{question}
Before proceeding with the answer, let us pause to understand the implication of this question. This question is another way of asking whether there is a way by which we can, by looking at rank of the (unification) gauge group of a single point in $\mathcal{C}$, determine the gauge groups at that point, after a putative symmetry breaking.
\par
The answer is, surprisingly, that the problem is NP-complete because of the following classic theorem in computational complexity.
\begin{thm}[Subset-sum problem, \cite{complexitybook}]\label{thm:subset}
Let $S$ be a set of positive integers and $S'$ a subset of $S$. Let $k$ be a fixed integer. Then the decision problem of deciding whether the sum of elements of $S'$ equals $k$ or not is NP-complete.
\end{thm}
It is intuitively clear that theorem \ref{thm:subset} answers question \ref{q:gauge}. However, a full rigorous proof has to attend to certain subtleties. Implicitly in our invoking the the subset-sum problem, we have made an assumption that the rank of a product of groups equals the sum of ranks of terms in that product. This is \emph{prima facie} only true when the product is the free product $\ast$ of groups (``the Grishko--H. Neumann theorem"):
$$\mathop{\rm rk}(G_1 \ast G_2) = \mathop{\rm rk} G_1 + \mathop{\rm rk} G_2.$$
For our purposes we have the same holding true because the free product descends to the tensor product at the Lie algebra level (for Lie algebras associated to the Lie groups in question) through its universal property.
\par
An interesting alternative to grand unification has been recently proposed by Donoghue and Pais \cite{donoghuepais}. In their work, instead of grand \emph{unification}, they propose a \emph{federation} of gauge groups by sequentially adding $SU(N)$ factors to $G_{SM}$ for large values of $N$ and such that the couplings converge at large enough energies. The hope then is that instead of unification, we should seek a fundamental explanation for $SU(N)$ gauge theories. At low energies, the gauge groups are ``autonomous".
\par
It is easy to see that even in this case, and perhaps in a more straight-forward way, that given a fixed gauge group rank ${\bf r}$, the problem of determing the ``constituent" gauge groups at low energies by rank considerations is also NP-complete, like question \ref{q:gauge} above. In this case for a sequence of ``independent" gauge groups $G_1, G_2, \ldots, G_n$ and $G_{SM}$ (fixed), we construct the free product
$$\Big( G_1 \ast G_2 \ast \cdots \ast G_n \Big) \ast G_{SM}$$ 
and basically repeat the argument as outlined above almost verbatim.

\section{Moduli, computation and fractals}
For a long time it was believed that the string Landscape-- the set of string vacua with small positive cosmological constant and containing the MSSM as a low-energy effective theory-- was infinite. This changed in 2006 with a paper by Acharya and Douglas \cite{acharyadouglas}. In this work it was argued that the string Landscape may be a ``discretum"; the authors argued their case on the basis of several deep ``finiteness" theorems in differential geometry and topology, including some results related to the geometrization conjecture. More specifically, the gist of the Acharya--Douglas argument was that since supergravity should be a manifestation of M-theory, one ought to look at relevant finiteness results in Riemannian geometry.
\par
One of the differential geometry results invoked was due to Cheeger which showed that, in a given sequence of smooth Riemannian manifolds $M_i$ with volumes, diameters and sectional curvature bounded, there can only be finitely many diffeomorphism types in $M_i$. This helped them bound the Kaluza-Klein compactification tower and the compactification volumes. Furthermore, they also invoked a theorem due to Gromov about the space of Riemannian manifolds with fixed dimension and bounded Ricci scalar and diameter being precompact in the Gromov-Hausdorff metric (more on that soon!). Acharya and Douglas argued that such convergence conditions on the space of manifolds were needed to support a central conjecture on the Landscape:
\begin{conj}[Acharya--Douglas \cite{acharyadouglas}, ``Hypothesis 1"]
There exists a minimal distance $\epsilon$ in the configuration space $\mathcal{C}$ between physically distinct vacua.
\end{conj}
I shall now argue, based on computation-theoretic grounds, that the problem of determining whether there is a ``minimal distance" between two points in $\mathcal{C}$ may be undecidable. In order to argue my case, let us take a detour through a fascinating world where geometry, topology and logic meet.
\par
A central (and rather intuitive!) problem in topology is deciding whether there exists an algorithm that tells, given two smooth manifolds $M$ and $N$, whether $M$ and $N$ are diffeomorphic (denoted as $M \simeq_{\mathop{\rm diff}} N$) or not. The general strategy for answering these types of questions algorithmically is to convert the problem into a decision questions about groups (fundamental groups, homology, \ldots). A rather famous theorem with this flavor is due to S. Novikov.
\begin{thm}[Novikov]\label{thm:novikov}
For all $n \geq 5$ and for all $n$-dimensional manifolds $M$ and a given fixed $n$-dimensional manifold $P$, it is unsolvable whether $P \simeq_{\mathop{\rm diff}}M$.
\end{thm}
One could ask a weaker question than that answered by theorem \ref{thm:novikov}, namely when is a homology $n$-sphere (with $n >5$ and sectional curvature bounded) diffeomorphic to the the standard $n$-sphere? Surprisingly enough, this problem-- intimately related to the Poincar\'e conjecture-- can be formulated as a halting problem. This was achieved by Nabutovsky and Weinberger \cite{weinnabu} who showed that, for a Turing degree of unsolvability $e \in \omega$ and for every Turing machine $T_e$, there is a sequence of homology $n$-spheres $\{P^e_k\}$, $k \in \omega$ such that $P^e_k \simeq_{\mathop{\rm diff}}S^n$ if and only if $T_e$ halts on input $k$ and \emph{the connected sum $N^e_k = P^e_k \# M \simeq_{\mathop{\rm diff}} M$ and $N^e_k$ is associated to a local minima of the diameter functional on the space $\mathop{\rm Met}(M)$ (the space of Riemmanian metrics on $M$ upto diffeomorphisms) with the depth of the local minima roughly equal to the settling time $\sigma_e(k)$ of the algorithm for input $y<k$} \cite{soare}. A slightly more formal statement is

\begin{thm}[``informal thm. 0.1" of \cite{weinnabu}]\label{thm:fractal}
For every closed smooth manifold of dimension $n > 4$, there are infinitely many local minima of the diameter functional $\mathop{\rm diam}: M \rightarrow \mathbb{R}$ on the subset $\mathop{\rm Al}(M) \subset \mathop{\rm Met}(M)$ of isometry classes of Riemannian metrics of curvature bounded in absolute value by 1 and the local minima is given by Riemannian metrics of smoothness $C^{1,\alpha}$ for $\alpha \in [0,1)$. Let $\beta$ be a computationally enumerable (c.e.) degree of unsolvability. Then there exists a positive constant $c(n)$ such that the local mimina of depth atleast $\beta$ is $\beta$-dense in a path metric on $\mathop{\rm Al}(M)$ and the number of $\beta$-deep minima where the diameter is always $\leq d$ is no less than $\exp(c(n)d^n)$.
\end{thm}

Theorem \ref{thm:fractal} may seem rather heavy-handed, just the sort of thing that mathematicians and nobody else would care about. Let us unpack it, focussing on the space
\begin{equation}\label{eq:Met}
\mathop{\rm Met}(M) := \mathop{\rm Riem}(M)/\mathop{\rm Diff}(M),
\end{equation}
to see its relevance for the study of string vacua\footnote{Denef and Douglas were the first to hint that the results of Nabutovsky and Weinberger could be of relevance to the Landscape. Unfortunately since the sequel of \cite{denefdouglas} is still awaited, we don't know what they would have done with it.}.
\par
First of all, we work in $\mathop{\rm Met}(M)$ as oppossed to $\mathop{\rm Riem}(M)$ directly since we want to impose a Gromov-Hausdorff (GH) metric on $\mathop{\rm Al}(M)$; roughly speaking the GH metric measures distances between metric spaces and we are, after all, interested in comparing how far two Riemannian metrics are from one another. Second of all, the theorem \ref{thm:fractal} tells us that the infinitely many local minima of the diameter functional are metrics of smoothness $C^{1, \alpha}$. These metrics have the same properties as Riemannian structures with sectional curvature between -1 and 1 (p.6 of \cite{weinnabu}) and therefore of obvious physical significance. The theorem also asserts that the number of $\beta$-deep minima is an exponential of a polynomial in the diameter bound of degree = dimension of the manifold. 
\par
In fact Nabutovsky and Weinberger prove more in \cite{weinnabu}. They show that the depth of the local minima (which is roughly the settling time of the algorithm, as noted above) is in fact of the order of the ``busy beaver function" in the dimension $n$. (The busy beaver function is an example of a function that grows faster than any computable function.) The unsolvability of the halting problem implies that \emph{one would never be able to determine if the depth has actually been computed} (p. 7 of \cite{weinnabu}).  One may also ask how farther apart are these basins. It is shown in \cite{weinnabu} that there are infinitely many deep basins between any two $\beta_1$- and $\beta_2$-deep minima for arbitrary Turing degrees $\beta_1$ and $\beta_2$. 
\par
We summarize the properties of the space $\mathop{\rm Met}(M)$ (\cite{weinnabu}, \cite{weinnabu2}, \cite{nabutovsygravity}) in the table below. (Notation: $\mathop{\rm inj}$ is the injectivity radius, $\mathop{\rm scal}$ denotes the Ricci scalar and the Einstein-Hilbert functional is defined w.r.t the standard volume form.)
\\
\begin{center}
  \begin{tabular}{| l || c | }
    \hline
                  & Riemannian metrics on smooth $M$  \\ \hline
      Functional & $\mathop{\rm vol}/\mathop{\rm inj^n}$ \\
    Moduli space & $\mathop{\rm Riem}(M)/{\mathop{\rm Diff}(M)}$\\ 
Local minima & Metrics of smoothness $C^{1,\alpha}, \alpha \in [0,1)$\\
Depth & $\exp(c(n)d^n)$ \\
Einstein-Hilbert functional & $\frac{\int_M \mathop{\rm scal}d\mu_M}{{\mathop{\rm vol}(M)}^{\frac{n-2}{n}}}$ \\
EH minimized (yes/no) & No (yes, after regularizing by adding $\epsilon |K| \mathop{\rm diam^2}$, $\epsilon > 0$)\\
Partition function & York--Gibbons--Hawking\\
Large scale structure & Fractal \\ \hline
\end{tabular}
\end{center}
\vspace{2mm}
In the Calabi Yau case, we are concerned with the properties of the space
\begin{equation}\label{eq:CYmod}
{\mathop{\rm Met}}_{J} (X)
\end{equation}
of metrics on a Calabi-Yau (CY) manifold $X$ with a fixed K\"ahler form $J$. Let us start with the classical definition of $X$. A CY manifold is a complex K\"ahler manifold with a trivial first Chern class and with a finite fundamental group \cite{yaudfn}. The space (\ref{eq:CYmod}) is the space of metrics on the CY manifold $X$. Now a natural question is what sort of metrics can be put on $X$? A deep theorem of Yau (settling Calabi's conjecture) is
\begin{thm}[Yau, cf. \cite{greenereview}]\label{thm:CY}
If $X$ is a complex K\"ahler manifold with vanishing first Chern class and with K\"ahler form $J$, then there exists a unique Ricci-flat metric on $X$ whose K\"ahler form $J'$ is in the same cohomology class as $J$.
\end{thm}
The existence of these metrics are of great importance in string phenomenology-- a Ricci-flat metric can be used to reduce the Einstein equation for gravity into a real Monge-Ampere equation and the solutions to the Einstein equation on a given CY can then describe the cosmological evolution of each point on the Landscape (or a universe in the multiverse.)
\par
Unfortunately, we know very few \emph{explicit} examples of Ricci-flat metrics on a given CY except in the most simple cases. On the other hand, we do have approximations to Ricci-flat metrics on a CY manifold to varying degrees of accuracy\footnote{Of course, we have, by definition, a family of Hermitian metrics on $X$!}. For example, the algebraic metrics recently studied in Headrick and Nassar \cite{headricknassar} seem to approximate any \emph{smooth} Ricci-flat metric to exponential accuracy. Furthermore, since the algebraic metrics are polynomials in the moduli parameters, several computational issues intrinsic to the solution of nonlinear PDEs such as the Einstein equations seem tractable in this approximation. The method of \cite{headricknassar} is to construct an energy functional out of the K\"ahler volume form in such a way that minimizing this functional yields the real Monge-Ampere equation. Together with the algebraic metrics, this yields a way to solve the Einstein equations on an \emph{algebraic} CY manifold.
\par
However, in presence of conifold singularities, the approximation of Ricci-flat metrics by algebraic metrics seem to be not so good. Another demand made in \cite{headricknassar} is that the manifolds in question should be geometrically uniform (that is, without large characteristic scales.) This raises the following question:
\begin{question}
Let $X$ be an arbitrary CY manifold (not geometrically uniform, with conifold singularities, \ldots). Describe the space (\ref{eq:CYmod}).
\end{question}
We are interested in setting up the problem in such way as to investigate the properties of the space (\ref{eq:CYmod}) in analogy to the space (\ref{eq:Met}). Let us start with some common points of similarity between the spaces (\ref{eq:Met}) and (\ref{eq:CYmod}).
\begin{enumerate}
\item{Convergence in Gromov-Hausdorff topology: in the Calabi-Yau case, we have, outside a singular set, every family of Ricci-flat metrics converging to a unique singular Ricci-flat metric in the Gromov-Hausdorff topology \cite{ruanzhang}.}
\item{Diameter bound: let $(X, \omega_0)$ be a compact $n$-dimensional Ricci-flat K\"ahler manifold and $\omega$ another Ricci-flat metric such that $\int_X \omega_0^{n-1} \wedge \omega \leq c_1$. Then the diameter of $(X, \omega_0)$ is bounded by $c_1, n$ and $\omega_0$ \cite{tosatti}\footnote{Compare this with the diameter bound in \cite{weinnabu}.}.}
\item{The CY analogue of $\mathop{\rm Al}(M)$: in the Riemannian case, we are concerned with path-metrics on $\mathop{\rm Al}(M)$ as defined above. Essentially, this is the space of metrics with sectional curvature bounded. The analogue of this in the CY cases exists through \cite{tosatti}: it is the space of all paths $\alpha_t: [0,1] \rightarrow \overline{\mathcal{K}_{NS}}$ where $\overline{\mathcal{K}_{NS}}$ is the closure of the ample cone.}
\end{enumerate}
\par
As remarked earlier, the problem of explicitly finding a Ricci-flat metric on a general Calabi-Yau is computationally very difficult. These facts coupled with the similarities between the spaces (\ref{eq:Met}) and (\ref{eq:CYmod}) suggest a conjecture similar to theorem \ref{thm:fractal}:
\begin{conj}\label{conj:main}
There exists a unique functional on the space (\ref{eq:Met}) such that the local minima is $\beta$-deep for $\beta$ a c.e. degree of unsolvability. Furthermore, the local minima are given by Ricci-flat K\"ahler metrics and parametrized by varying K\"ahler and complex moduli. The number of $\beta$-deep local minima with diameter bounds by $c_1$ and $n$ (for a fixed $\omega_0$) is an exponential of a polynomial in $c_1$ and $n$.
\end{conj}
Evidently, if this conjecture is true, then we would have a conceptual explanation as to why it is so hard to find an explicit Ricci-flat metric in a given K\"ahler class. It would, of course, also imply that the problem of exactly solving the Einstein equations on an arbitrary Calabi-Yau manifold with an explicit Ricci-flat metric is computationally and conceptually much harder than anticipated before.

\section{Periods and string theory vacua}
The last point that I'd like to bring forth about questions of algorithmic decidability and the Landscape concerns fundamental \emph{periods} of Calabi-Yau manifolds. It is a basic claim \cite{berglund} that the low energy effective theory with $N=2$ supersymmetry on a Calabi-Yau 3-fold $X$ (both the Yukawa couplings and the K\"ahler potential) is encoded in the periods of the manifold which is defined as
\begin{equation}\label{eq:period}
\varpi_i := \int_{\gamma^{i}} \Omega,
\end{equation}
where $\Omega$ is a holomorphic 3-form and $\gamma^{i}$ the basis of homology cycles of $X$. The fundamental period of $X$ is $\varpi_0$. It is shown in \cite{berglund} that $\varpi_0$ can be explicitly computed for a large class of Calabi-Yau (such as those realized as hypersurfaces in the weighed projective space or of the complete intersection type). For example, for a one parameter family of mirrors of quintic 3-folds $M/G$ with $M$ given by the zero locus of the polynomial $p(x, \psi) = \sum_{k=1}^5 x_k^5 - 5\psi x_1 \cdots x_5$ and the coordinates of $M$ identified under the action of $G= \mathbb{Z}_5^3$, it can be shown that
\begin{equation}\label{eq:basic}
\varpi_0(\psi) = \sum_{n=0}^\infty \frac{(5n)!}{{(n!)}^5{(5 \psi)}^{5m}}
\end{equation}
where $|\psi| \geq 1$ and $0 < \arg(\psi) < \frac{2\pi}{5}$. After analytic continuation to $|\psi| < 1$, we get from (\ref{eq:basic})
\begin{equation}\label{eq:analcon}
\varpi_0(\psi) = - \frac{1}{5} \sum_{m=1}^\infty \frac{\Gamma\Big(\frac{m}{5}\Big) {(5 \alpha^2 \psi)}^m}{\Gamma(m) \Gamma^4\Big(1- \frac{m}{5}\Big)}.
\end{equation}
It is a theorem that for low-energy N=2 SUSY effective theory, the periods of Calabi-Yau hypersurfaces in weighed projective space with more than two moduli parameters can be expressed in terms of iterated Mellin-Barnes integrals and Horn series \cite{passare}.
\par
Periods, of course, are a basic arithmetic object. While periods in superstring theory are really \emph{functions} of certain parameters (say of $\psi$ in (\ref{eq:analcon})), periods in arithmetic algebraic geometry are \emph{numbers} of a very specific form ``lying" between the algebraic closure $\overline{\mathbb{Q}}$ and $\mathbb{C}$. They are obtained from integrating an algebraic differential form over a cycle in an algebraic variety (generally defined over $\mathbb{Q}$) \cite{kz}. It can be verified that in the string-theoretic setting of Calabi-Yau manifolds realized as hypersurfaces in weighed projective spaces, we obtain periods in this number-theoretic sense for all values of the multiple moduli parameters.
\par
It is, in general, a very difficult question determining whether (1) a given number is a period or not and (2) verifying whether two periods are the same or not. In the paper \cite{kz}, Kontsevich and Zagier give several nontrivial equalities between periods. They, furthermore, conjecture that one period can be expressed as another through three basic algebraic operations-- (1) additivity, (2) change of variables and (3) the Stokes formula.
\par
One idiosyncratic view of the string Landscape (and this is decidely mine alone!) is imagining the entire configuration space of vacua $\mathcal{C}$ as a collection of fundamental periods, one for each compactification. We can then ask whether two points on $\mathcal{C}$ are the same or not by asking whether the corresponding two fundamental periods are the same or not, for random choices of the parameter values. Notice that this is a highly relevant question for phenomenology since the merits of periods lie in their ``knowledge" of the low-energy effective theory.  
\par
Now as Kontsevich and Zagier remark, in general, this question is likely to be ``completely intractable now and may remain so for many years" (p.8 of \cite{kz}). The question is whether there is any precise way by which we can quantify this intractability. Essentially, from the computational complexity point of view, we want to get a better understanding of 
\begin{enumerate}
\item{What sort of numbers are periods, from the computation-theoretic viewpoint?}
\item{How can we distinguish periods based on their complexity such that for two given periods, it would be ``easy" (= doable in polynomial time) to check whether the two are the same or not if and only if they are in the same complexity class?}
\end{enumerate}
Question 1 has been already answered by Yoshinaga \cite{yoshinaga}: in a very interesting paper, he shows that all (real) periods are computable in the sense of Turing. The proof is a mixture of facts from Tarski's quantifier-elimination theory and semi-analytic geometry of Hironaka. Let me make some brief remarks about how to tackle question 2. Let $I(s)$ be the \emph{Igusa zeta function}
\begin{equation}\label{eq:izeta}
I(s) = \int_{\Delta_n} f^s \omega,
\end{equation}
where $s$ is a complex variable, $\omega = dx_1 \wedge \cdots \wedge dx_n$ and $\Delta_n \subset \mathbb{R}^{n+1}$ be the standard simplex with volume form $\omega$. By a theorem of Belkale--Brosnan \cite{belkale}, we know that if $f$ is a polynomial in $N$-variables and with $\mathbb{Q}$-coefficients and $s_0$ an integer, then the Laurent expansion of the Igusa zeta function 
\begin{equation}\label{eq:laurent}
I(s) = \sum_{i > N} a_i {(s-s_0)}^i
\end{equation}
has coefficients $a_i$ which are periods. Now we can associate to Igusa zeta functions a canonical measure of complexity of functions, namely ``heights". The idea, very roughly, for the construction of complexity classes for periods would be to compare the heights so associated to the periods $a_i$ (cf. \cite{chambertloir} for the relationship between heights and the Igusa zeta function).

\section{Conclusion}
In this paper, I have put forward three areas through which questions of decidability and computational complexity enter the discussion of the string-theoretic Landscape: (1) through the determination of the average gauge group, (2) the construction of explicit Ricci-flat metrics on a general Calabi-Yau manifold and finally (3) through the comparison of fundamental periods of Calabi-Yau compactifications which give a low-energy effective $N=2$ SUSY theory. In conjunction with the results of Denef-Douglas on the cosmological constant, these ideas present a fairly grim picture of what we \emph{can} know about the Landscape. Of course, the ideas presented here are the ``worst-case" scenarios in the sense that we have highlighted areas where computation-theoretic ideas \emph{could} give rise to undecidable questions. (Most of the assertions here are, the reader is reminded, conjectural.) Nevertheless, it is my belief that computation theory will play a larger role in any description of the Landscape in the years to come. If this is the case, it will certainly validate a often-repeated claim (cf. \cite{tegmark}) of the role of algorithmic decidablity in an ``ultimate" theory of the material universe.

\bibliographystyle{amsalpha}

\begin{thebibliography}{A}

\bibitem[1]{acharyadouglas} Bobby Acharya, Michael Douglas \textit{A finite Landscape?} arXiv:hep-th/0606212v1.

\bibitem[2]{belkale} Prakash Belkale, Patrick Brosnan \textit{Periods and Igusa zeta functions} arXiv:math/0302090v1 [math.NT].

\bibitem[3]{berglund} Per Berglund, Philip Candelas, Xenia de la Ossa, Anamarie Font, Tristan Hubsch, Dubravka Jancic and Fernando Quevedo \textit{Periods for Calabi--Yau and Landau--Ginzburg vacua} Nucl. Phys. B \textbf{419} (1994) 352--403.

\bibitem[4]{blumenhagen} Ralph Blumenhagen, Florian Gmeiner, Gabriele Honecker, Dieter L\"ust and Timo Weigland \textit{The statistics of supersymmetric D-brane models} Nucl. Phys. B \textbf{713} (2005) 83--115.

\bibitem[5]{chambertloir} Antoine Chambert-Loir \textit{Lectures on height functions} arXiv: 0812.0947v1 [math.NT].

\bibitem[6]{denefdouglas} Fredrik Denef, Michael R. Douglas \textit{Computational complexity of the landscape I} Annals Phys. \textbf{322} (2007) 1096--1142.

\bibitem[7]{distlergaribaldi} Jacques Distler, Skip Garibaldi \textit{There is no ``Theory of Everything" inside E8} arXiv: 0905.2658v2 [math.RT].

\bibitem[8]{donoghuepais} John F. Donoghue, Preema Pais \textit{Gauge federation as an alternative to unification} Phys. Rev. D \textbf{79} (2009) 095020.

\bibitem[9]{complexitybook} Michael R. Garey, David S. Johnson \textit{Computers and Intractability: A Guide to the Theory of NP-Completeness}, W.H. Freeman, 1979.

\bibitem[10]{gmeiner} Florian, Gmeiner, Ralph Blumenhagen, Gabriele Honecker, Dieter L\"ust, Timo Wiegland \textit{One in a Billion: MSSM-like D-Brane Statistics} JHEP \textbf{0601} (2006) 004. 

\bibitem[11]{greenereview} Brian R. Greene \textit{String theory on Calabi-Yau manifolds} arXiv:hep-th/9702155v1.

\bibitem[12]{headricknassar} Matthew Headrick, Ali Nassar \textit{Energy functionals for Calabi-Yau metrics} arXiv: 0908.2635v1 [hep-th]

\bibitem[13]{kz} Maxim Kontsevich, Don Zagier \textit{Periods} Mathematics unlimited---2001 and beyond, Springer Berlin (2001) 771--808.

\bibitem[14]{kumarwells} Jason Kumar, James D. Wells \textit{Landscape cartography: a coarse survey of gauge group rank and stabilization of the proton} Phys. Rev. D \textbf{71} (2005) 026009.

\bibitem[15]{nabutovsygravity} Alexander Nabutovsky \textit{Combinatorics of the space of Riemannian structures and logic phenomena of Euclidean quantum gravity} Perspectives in Riemannian geometry, CRM Proceedings and Lecture Notes 40, Amer. Math. Soc. Providence RI (2006) 223--248

\bibitem[16]{weinnabu2} Alexander Nabutovsky, Schmuel Weinberger \textit{Critical points of Riemannian functionals and arithmetic groups} Publ. IH\'ES \textbf{92} (2000) 5--62.

\bibitem[17]{weinnabu} Alexander Nabutovsky, Schmuel Weinberger \textit{The fractal nature of Riem/Diff I} Geometriae Dedicata \textbf{101} (2003) 145--250.

\bibitem[18]{passare} M. Passare, A.K. Tsikh, A.A. Cheshel \textit{Multiple Mellin-Barnes Integrals as Periods of Calabi-Yau Manifolds With Several Moduli} Theor.Math.Phys. \textbf{109} (1997) 1544--1555.

\bibitem[19]{rej} Abhijnan Rej \textit{Algorithms and the Landscape}  in preparation (2009).

\bibitem[20]{ruanzhang} Wei-Dong Ruan, Yuguang Zhang \textit{Convergence of Calabi-Yau manifolds} arXiv: 0905.3424v1 [math.DG]

\bibitem[21]{smolin} Lee Smolin \textit{The trouble with physics: the rise of string theory, the fall of a science and what comes next} Penguin Books London (2008)
 
\bibitem[22]{soare} Robert I. Soare \textit{Computability theory and differential geometry} Bull. Symbolic Logic (2004)

\bibitem[23]{tegmark} Max Tegmark \textit{The mathematical universe} Found. Phys. \textbf{38} (2008) 101-150.

\bibitem[24]{tosatti} Valentino Tosatti \textit{Limits of Calabi-Yau metrics when the K\"ahler class degenerates} arXiv: 0710.4579v1 [math.DG].

\bibitem[25]{vilenkin} Alexander Vilenkin \textit{Perspectives in cosmology} arXiv:0908.0721v1 [astro-ph.CO].

\bibitem[26]{yaudfn} Shing-Tung Yau \textit{Calabi-Yau manifolds} Scholarpedia \textbf{4(8):6524} (2009).

\bibitem[27]{yoshinaga} Masahiko Yoshinaga \textit{Periods and elementary real numbers} arXiv:0805.0349v1 [math.AG].

\end{thebibliography}

\end{document}